# A hot Jupiter orbiting a 2-Myr-old solar-mass T Tauri star

JF Donati[1,2], C Moutou[3], L Malo[3], C Baruteau[1,2], L Yu[1,2], E Hébrard[4], G Hussain[5], S Alencar[6], F Ménard[7,8], J Bouvier[7,8], P Petit[1,2], M Takami[9], R Doyon[10], A Collier Cameron[11]

**Hot Jupiters are giant Jupiter-like exoplanets that orbit 100x closer to their host stars than Jupiter does to the Sun. These planets presumably form in the outer part of the primordial disc from which both the central star and surrounding planets are born, then migrate inwards and yet avoid falling into their host star[1]. It is however unclear whether this occurs early in the lives of hot Jupiters, when still embedded within protoplanetary discs[2], or later, once multiple planets are formed and interact[3]. Although numerous hot Jupiters were detected around mature Sun-like stars, their existence has not yet been firmly demonstrated for young stars[4,5,6], whose magnetic activity is so intense that it overshadows the radial velocity signal that close-in giant planets can induce. Here we show that hot Jupiters around young stars can be revealed from extended sets of high-resolution spectra. Once filtered-out from the activity, radial velocities of V830 Tau derived from new data collected in late 2015 exhibit a sine wave of period 4.93±0.05 d and semi-amplitude 75±11 m/s, detected with a false alarm probability <0.03%. We find that this signal is fully unrelated to the 2.741-d rotation period of V830 Tau and we attribute it to the presence of a 0.77±0.15 $M_{2\downarrow}$ planet orbiting at a distance of 0.057±0.001 au from the host star. Our result demonstrates that hot Jupiters can migrate inwards in <2 Myr, most likely as a result of planet-disc interactions, and thus yields strong support to the theory of giant planet migration in gaseous protoplanetary discs[2].**

Very few exoplanets have yet been discovered around young forming Sun-like stars aged <10 Myr[7,8], the so-called T Tauri stars (TTSs), either through the radial velocity (RV) variations or the photometric transits they induce in the light of their stars. Yet detections of young planets are key for our understanding of how planetary systems form; this is especially true of young hot Jupiters (hJs) thought to have a critical impact on the early architecture of these systems. The first claimed detection of a young hJ[4] orbiting a TTS was quickly refuted; the reported periodic RV fluctuations were finally attributed to activity[5] and to cool spots at the stellar surface[9]. The recent candidate detection of a transiting hJ around a TTS[6] is still pending confirmation.

TTSs are known to harbor spots and plages at their surfaces, generating RV fluctuations with semi-amplitudes of several km/s[10], i.e., much larger than the perturbations expected from a putative planet, even for close-in massive hJs inducing typical RV signals of ~0.1 km/s. Detecting hJs around TTSs through velocimetry or photometry is thus quite challenging and mandatorily requires efficient tools for filtering-out the dominant jitter that activity induces in the spectra and light-curves of young stars. We recently proposed a new method to achieve this goal[11], whose first applications to wTTSs proved promising though inconclusive[12].

V830 Tau is a ~2-Myr-old solar-mass TTS[12] contracting towards the main sequence and currently spinning in 2.741 d[13], i.e., ~10x faster than the Sun. Evolutionary models[14] suggest that it is fully or largely convective. Unlike 80% of the TTSs in the Taurus star-forming region[15], V830 Tau exhibits no significant infrared excess, implying that most of its inner accretion disc already

[1]Univ. de Toulouse, UPS-OMP, IRAP, 14 av Belin, F–31400 Toulouse, France. [2]CNRS, IRAP / UMR 5277, 14 av Belin, F–31400 Toulouse, France. [3]CFHT Corporation, 65-1238 Mamalahoa Hwy, Kamuela, Hawaii 96743, USA. [4]Physics & Astronomy, York University, Toronto, Ontario L3T 3R1, Canada. [5]ESO, Karl-Schwarzschild-Str. 2, D-85748 Garching, Germany. [6]Departamento de Fisica, ICEx, UFMG, av Antonio Carlos, 6627, 30270-901 Belo Horizonte, MG, Brazil. [7]Univ. Grenoble Alpes, IPAG, BP 53, F–38041 Grenoble Cédex 09, France. [8]CNRS, IPAG / UMR 5274, BP 53, F–38041 Grenoble Cédex 09, France. [9]Institute of Astronomy & Astrophysics, Academia Sinica, PO Box 23-141, 106, Taipei, Taiwan. [10]Dép. de physique, Univ. de Montréal, CP 6128, Succ. Centre-Ville, Montréal, QC, Canada H3C 3J7. [11]SUPA, School of Physics and Astronomy, Univ. of St Andrews, St Andrews, Scotland KY16 9SS, UK

dissipated. This is consistent with its status as a non-accreting weak-line TTS (wTTS), and makes it an ideal target to look for the presence of hJs at an early stage of star and planet formation.

In late 2015, we collected 48 high-resolution spectra of V830 Tau (see Extended Table 1) as part of the MaTYSSE Large Program aimed at detecting hJs around wTTSs[11]. Applying least-squares deconvolution[16] (LSD) to our spectra, we derived accurate average line profiles and their temporal modulation over ~15 rotation cycles. Longitudinal magnetic fields were also derived from our circularly-polarized data and the Zeeman signatures that fields generate in spectral lines[16]. Using tomographic techniques inspired from medical imaging, one can reconstruct distributions of spots and plages at the surfaces of rotating cool active stars from sets of densely-sampled line profiles covering several rotation cycles. This method, called Doppler imaging[17] (DI), can also probe the photospheric shear associated with surface differential rotation through the amount of twisting it generates in brightness maps[18,19].

Our DI code was previously applied to a small set of 15 LSD profiles of V830 Tau, from which the distribution of surface features and the differential rotation pattern were recovered[12]; it even suggested the potential presence of a hJ, though with a very low confidence level. The version presented herein implements a novel technique to filter-out lunar contamination (plaguing spectra collected in non-photometric conditions) yielding good results when phase coverage is dense. Applying the code to our new set of 48 LSD profiles of V830 Tau yields the map and fit shown in Fig 1. We again clearly detect differential rotation (see Extended Fig 1) and confirm that it is ~3x weaker than that of the Sun, reflecting that V830 Tau is largely or fully convective[20].

From the brightness image reconstructed with DI, we derive the model RV curve that V830 Tau should exhibit if all profile perturbations were attributable to surface features and differential rotation. By subtracting these modeled RVs from the observed ones (both computed as the first moment of the LSD profiles), we obtain the activity-filtered RVs of V830 Tau whose amplitude is typically ~10x lower than the raw RVs (see Fig 2). A clear RV signal, with a period of 4.93±0.05 d and a semi-amplitude of 75±11 m/s, is detected in the activity-filtered RVs at a confidence level >99.97% (see Figs 2 & 3). Using a Bayesian approach[21], we find that the 4.93-d peak is at least $10^5$ times more likely than any of the other features in the periodogram (see Extended Fig 2a). The regular phase coverage of our data also allows us to check that the 4.93-d signal is present in smaller subsets (e.g., first and second half, even and odd points) though expectedly with a lower confidence level; we similarly checked that our detection holds when profiles affected by lunar contamination are excluded (see Fig 3 and Extended Fig 2a) and when differential rotation is neglected. Periodograms of the longitudinal fields and of the Hα emission, both reliable proxies for the activity jitter plaguing RV curves[22], show no power at a period of 4.93 d (see Fig 3b & Extended Fig 2b), demonstrating that the signal we report is unrelated to activity.

We interpret this RV signal as caused by a giant planet of mass 0.77±0.15 $M_{2|}$ in circular orbit around V830 Tau at a distance of 0.057±0.001 au (see Extended Table 2). Although the filtered RVs are marginally better fitted for an eccentric orbit (e=0.30±0.15, see Fig 2), we still favor a circular orbit given the large error bar on the eccentricity[23]. Removing the planet signal from the original data and repeating the activity-filtering process yields residual RVs with a rms dispersion of 48 m/s (i.e., consistent with the average noise level, see Extended Table 1) and no significant peak left in the periodogram. Using the alternative option of fitting both the brightness map and the planet parameters simultaneously[24] yields identical results and demonstrates that our optimal model including a planet is orders of magnitude more likely than a model with no planet (see Fig 4). Simulations in conditions identical to those of our observations yield results in close agreement with those of Figs 2 and 3 (see Extended Figs 3 & 4), further demonstrating that our filtering process induces no spurious peak and that the RV signal we detect is unrelated to activity.

A careful re-analysis of our original data[12] demonstrates that, despite being affected by intrinsic variability from the host star, they nonetheless confirm the presence of the planet signal detected in our new data. In particular, applying our filtering analysis to the main subset of our original data (featuring dense enough coverage for our technique to perform reliably) yields filtered RVs agreeing well with those derived from the new data; fitting both sets together further improves the confidence level at which the planet is detected (see Extended Fig 2c).

The detection we report among the small sample of wTTSs (~10) already studied with MaTYSSE suggests that close-in giant planets are potentially more frequent around TTSs than around mature low-mass stars, ~1% of which are known to host hJs[25,26]. Poisson statistics however indicate that there is still a ~10% chance that hJs are similarly frequent for both populations. A more quantitative conclusion will have to await for a thorough analysis of all MaTYSSE data, later-on complemented by the large TTS survey to be carried out with SPIRou, the new nIR cryogenic spectropolarimeter and velocimeter to be installed at CFHT in 2017.

The broad consensus is that hJs form beyond a few au from their host stars and migrate inwards to their close-in orbits. The delivery of hJs may result from planet-disc interactions[2] (disc migration) or from dynamical interactions with a planetary[3] or a stellar companion, followed by orbital circularization (high-eccentricity migration). These highly-debated migration channels were proposed as tentative explanations for hJs with orbits either well aligned or misaligned with the spin axis of their host stars[27]. Our detection of a hJ on a ~5-d circular (or moderately eccentric) orbit around a 2-Myr-old star is most naturally explained with hJ delivery by disc migration, producing hJs with nearly circular orbits, rather than by planet-planet scattering, generating highly-eccentric (e>0.9) hJs whose circularization timescales are at least 2-3 orders of magnitude longer[28] than the age of V830 Tau (for typical tidal dissipation factors in giant planets[29]). Our result thus yields strong support to the theory of Type-II migration of giant planets in gaseous protoplanetary discs[2] and confirms that the architecture of planetary systems, on which hJs have a strong impact, is likely quite dynamic right from the very early ages of planetary formation.

Global models of planet formation and evolution show that giant planets can reach a mass and an orbital period similar to those of V830 Tau b in 2-3 Myr, whether formation occurs through core accretion or disk gravitational instability. Large uncertainties in these models make it difficult to accurately predict the occurrence rate of hJs and thus to associate V830 Tau b to either formation scenario. The ~300 G dipole of the star's magnetic field[12] is strong enough to have disrupted the central 0.06 au of the now-dissipated disc for accretion rates $<2.10^{-10}$ $M_\odot$/yr; for rates of ~$10^{-9}$ $M_\odot$/yr, more typical to those of the classical TTSs (cTTSs) that still feed from their discs, a >700 G dipole field is required, again compatible with observations of cTTSs similar to V830 Tau[30,12]. This shows that the field of V830 Tau may well have stopped the planet within the magnetospheric gap[1] at the end of its disc migration, and saved it from falling into the star.

**References**

1. Lin, D.N.C., Bodenheimer, P., Richardson, D.C. Orbital migration of the planetary companion of 51 Pegasi to its present location. *Nature* **380**, 606-607 (1996)
2. Baruteau, C., et al. Planet-Disk Interactions and Early Evolution of Planetary Systems. Protostars and Planets VI, *Univ. Arizona Press* **914**, 667-689 (2014)
3. Chatterjee, S., Ford E.B., Matsumura, S., Rasio, F.A. Dynamical Outcomes of Planet-Planet Scattering. *Astrophys. J.* **686**, 580-602 (2008)
4. Setiawan, J., et al. A young massive planet in a star-disk system. *Nature* **451**, 38-41 (2008)
5. Huelamo, N., et al. TW Hydrae: evidence of stellar spots instead of a Hot Jupiter. *Astron. & Astrophys.* **489**, L9-L13 (2008)
6. van Eyken, J.C., et al. The PTF Orion project: a possible planet transiting a T-Tauri star. *Astrophys. J.* **755**, 42-55 (2012)
7. Kraus, A.L., Ireland, M.J. LkCa 15: A Young Exoplanet Caught at Formation? *Astrophys. J.* **745**, 5-16 (2012)
8. Sallum, S., et al. Accreting Protoplanets in the LkCa 15 Transition Disc. *Nature* **527**, 342-344 (2015)
9. Donati, J.-F., et al. The large-scale magnetic field and poleward mass accretion of the classical T Tauri star TW Hya et al. *Mon. Not. R. Astron. Soc.* **417**, 472-487 (2011)

10. Mahmud, N.I., et al. Starspot-induced Optical and Infrared Radial Velocity Variability in T Tauri Star Hubble I 4. *Astrophys. J.* **736**, 123-131 (2011)
11. Donati, J.-F., et al. Modeling the magnetic activity and filtering radial velocity curves of young Suns : the weak-line T Tauri star LkCa 4. *Mon. Not. R. Astron. Soc.* **444**, 3220-3229 (2014)
12. Donati, J.-F., et al. Magnetic activity and hot Jupiters of young Suns: the weak-line T Tauri stars V819 Tau and V830 Tau. *Mon. Not. R. Astron. Soc.* **453**, 3706-3719 (2015)
13. Grankin, K.N., Bouvier, J., Herbst, W., Melnikov, S.Y. Results of the ROTOR-program. II. The long-term photometric variability of weak-line T Tauri stars. *Astron. & Astrophys.* **479**, 827-843 (2008)
14. Siess, L., Dufour, E., Forestini, M., An internet server for pre-main sequence tracks of low- and intermediate-mass stars. *Astron. & Astrophys.* **358**, 593-599 (2000)
15. Kraus, A.L., Ireland, M.J., Hillenbrand, L.A., Martinache, F. The Role of Multiplicity in Disk Evolution and Planet Formation. *Astrophys. J.* **745**, 19-29 (2012)
16. Donati, J.-F., Semel, M., Carter, B.D., Rees, D.E., Collier Cameron, A. Spectropolarimetric observations of active stars. *Mon. Not. R. Astron. Soc.* **291**, 658-682 (1997)
17. Vogt, S.S., Penrod, G.D., Hatzes, A.P. Doppler images of rotating stars using maximum entropy image reconstruction, *Astrophys. J.* **321**, 496-515 (1987)
18. Donati, J.-F., Collier Cameron, A. Differential rotation and magnetic polarity patterns on AB Doradus, *Mon. Not. R. Astron. Soc.* **291**, 1-19 (1997)
19. Donati, J.-F., Collier Cameron, A. Petit, P. Temporal fluctuations in the differential rotation of cool active stars. *Mon. Not. R. Astron. Soc.* **345**, 1187-1199 (2003)
20. Morin, J., et al. The stable magnetic field of the fully convective star V374 Peg. *Mon. Not. R. Astron. Soc.* **384**, 77-86 (2008)
21. Mortier, A., Faria, J.P., Correia, C.M., Santerne, A., Santos, N.C. BGLS: A Bayesian formalism for the generalised Lomb-Scargle periodogram. *Astron. & Astrophys.* **573**, 101-106 (2015)
22. Haywood, R.D., et al. The Sun as a planet-host star: Proxies from SDO images for HARPS radial-velocity variations. *Mon. Not. R. Astron. Soc.* **457**, 3637-3651 (2016)
23. Lucy, L.B., Sweeney, M.A. Spectroscopic binaries with circular orbits. *Astron. J.* **76**, 544-556 (1971)
24. Petit, P., et al. A maximum entropy approach to detect close-in giant planets around active stars. *Astron. & Astrophys.* **584**, 84-91 (2015)
25. Mayor, M., et al. The HARPS search for southern extra-solar planets. Occurrence, mass distribution & orbital properties of super-Earths and Neptune-mass planets. arXiv:1109.2497 (2011)
26. Wright, J.T., et al. The Frequency of Hot Jupiters Orbiting nearby Solar-type Stars. *Astrophys. J.* **753**, 160-164 (2012)
27. Winn, J.N., Fabrycky, D.C. The Occurrence and Architecture of Exoplanetary Systems. *Annual Rev. Astron. & Astrophys.* **53**, 409-447 (2015)
28. Ogilvie, G.I. Tidal Dissipation in Stars and Giant Planets. *Annual Rev. Astron. & Astrophys.* **52**, 171-210 (2014)
29. Goldreich, P., Soter, S. Q in the Solar System. *Icarus* **5**, 375-389 (1966)
30. Donati, J.-F., et al. Magnetometry of the cTTS GQ Lup: non-stationery dynamos and spin evolution of young Suns. *Mon. Not. R. Astron. Soc.* **425**, 2948-2963 (2012)

**Acknowledgements**

This paper is based on observations obtained at the CFHT (operated by the National Research Council of Canada, the Institut National des Sciences de l'Univers of the Centre National de la Recherche Scientifique of France and the University of Hawaii), at the TBL (operated by Observatoire Midi-Pyrénées and by the Institut National des Sciences de l'Univers of the Centre National de la Recherche Scientifique of France), and at the Gemini Observatory (operated by the Association of Universities for Research in Astronomy, Inc., under a cooperative agreement with the NSF on behalf of the Gemini partnership: the National Science Foundation of the United States, the National Research Council of Canada, CONICYT of Chile, Ministerio de Ciencia, Tecnología e Innovación Productiva of Argentina, and Ministério da Ciência, Tecnologia e Inovação of Brazil). We thank the QSO teams of CFHT, TBL and Gemini for their great work and efforts, without which this study would not have been possible. We also thank the IDEX initiative of Univ

Fédérale Toulouse Midi-Pyrénées for awarding a "Chaire d'Attractivité" to GH, in the framework of which this work was done. SA acknowledges financial support from CNPq, CAPES and Fapemig.

**Author Contributions** This work merged data collected with 2 different instruments and 3 different telescopes. JFD led the data processing, analysis and manuscript preparation, CM, LM, LY and EH participated in the data collection and data analysis, whereas CB contributed to the theoretical implications of the results. All coauthors, including GH, SA, MT, FM, JB, PP, RD and ACC, were involved in elaborating the observing proposals, in discussing the results at various stages of the analysis, and in providing contributions to older versions of the manuscript.

**Author Information** The authors declare no competing financial interests. Correspondence should be addressed to JFD (**jean-francois.donati@irap.omp.eu**)

## Main Figure Legends

**Figure 1 | Brightness map of V830 Tau and fit to the LSD profiles. a**, Logarithmic brightness at the surface of V830 Tau as derived with DI. Cool spots / bright plages show up as brown / blue features. The rotation axis of the star is tilted at 55° to the line of sight, and the projected equatorial rotation velocity is equal to 30.5 km/s[12]. The star is shown in a flattened polar view, with the pole in the centre and the equator depicted as a bold circle. Ticks outside the image mark phases of observations. **b**, Observed (black line) and modeled (red) LSD profiles of V830 Tau throughout our 2015 November-December run. LSD profiles prior to their filtering from lunar contamination are also shown (cyan). Numbers right to each profile indicate the rotation cycle. Cycles 6.079 to 7.190 are the most impacted, contamination being much smaller or negligible in all other observations.

**Figure 2 | Raw, filtered and residual RVs of V830 Tau. a,** Top curve: raw RVs of V830 Tau (open symbols and 1σ error bars) and model inferred with Doppler imaging (cyan line); the model slowly evolves with time as a result of differential rotation. Open circles, squares and triangles depict ESPaDOnS, NARVAL and ESPaDOnS / GRACES data. Middle curve: activity-filtered RVs and best sine fit to the data (cyan line). The period and semi-amplitude of the planet RV signal are equal to 4.93±0.05 d and 75±11 m/s (1σ error bars). Bottom curve: residual RVs once the planet signal is removed and activity is filtered out, with a final rms dispersion of 48 m/s. Colors code the rotation cycles. **b**, Activity-filtered RVs phase-folded on the planet orbital period of 4.93 d. Although the fit to the data is marginally better with an eccentric orbit (dashed line) than with a circular orbit (solid line), the significance of the derived eccentricity (0.30±0.15) is too low to be reliable[23].

**Figure 3 | Periodogram of the RVs and activity of V830 Tau. a,** Periodograms of the raw (top), filtered (middle) and residual RVs (bottom) shown in Fig 2a. The black line is for the full set, while the dashed red, green, blue and pink lines are for the first half, the second half, the even points and the odd points only; the purple (resp orange) dashed lines are the periodograms once the 15 RV points affected (resp the 6 RV points strongly affected) by lunar contamination are removed. (Similar results are obtained when analyzing subsets of filtered RVs from a global DI modeling, or when DI filtering is applied to subsets of LSD profiles). The stellar rotation period (2.741 d), its first harmonic and the planet orbital period (4.93 d) are depicted with vertical dashed lines. The horizontal dotted and dashed lines trace the 33%, 10%, 3% and 1% false alarm probabilities (FAP). The planet signal in the filtered RVs is detected in the full set with a FAP <0.03%. **b,** Periodogram of the line-of-sight projected (longitudinal) magnetic field, a reliable activity proxy[22], featuring a clear peak at the stellar rotation period but no power at the planet orbital period.

**Figure 4 | Adjusting the planet parameters while modeling activity.** Variations in the reduced $\chi^2$ of the DI fit to the LSD profiles for a given spottedness level and assuming the presence of a planet in circular orbit, for a range of orbital periods $P_{orb}$ and semi-amplitudes K of the RV signature. (This is a 2D cut from a 3D map, with the phase of the RV signal also included as a search parameter). The location of the minimum and local paraboloid curvature yield the optimal values of $P_{orb}$ and K and their respective 1$\sigma$ error bars[24], equal to 4.94±0.05 d and 82±10 m/s, in agreement with the results of our main filtering technique. The outer color contour traces the projected 99.99% confidence interval, corresponding to a $\chi^2$ increase of 21.1 for a 3-parameter fit to the 2208 data points of the LSD profiles. With respect to our best model incorporating a planet, a model with no planet corresponds to a $\chi^2$ increase of 82, implying that the planet is detected with a FAP <$10^{-15}$, much smaller than the FAP derived from the periodogram of the 48 RV points (see Fig 3) thanks to the larger number of data points in the fitted LSD profiles.

## Methods

### Spectropolarimetry with ESPaDOnS / NARVAL and the MaTYSSE programme

ESPaDOnS[31] and NARVAL are twin spectropolarimeters respectively installed at the Cassegrain focus of the 3.6m Canada-France-Hawaii Telescope (CFHT) atop Maunakea (Hawaii), and of the 2m Bernard Lyot Telescope atop Pic du Midi (France). Both include a fiber-fed bench-mounted high-resolution spectrograph, yielding full coverage of the 370–1,000 nm wavelength range in a single exposure at a spectral resolving power of 65,000. ESPaDOnS can also be fed from the 8m Gemini-N Telescope next to CFHT, through a 300m fiber link called GRACES[32], yielding spectra with either similar resolving power (in star-only mode) or half of it (in star+sky mode). In this run, we secured 16 spectra with ESPaDOnS from 2015 Nov 17 to Dec 02 in fair weather conditions, 16 spectra with NARVAL from 2015 Nov 10 to Dec 17 in moderate to good weather, and 16 spectra with ESPaDOnS / GRACES on 4 different nights towards the end of the run (with 2 / 14 spectra in star-only / star+sky mode respectively). The full journal of observations is given as Extended Table 1. ESPaDOnS & NARVAL observations were secured in spectropolarimetric mode (circular polarization) in the framework of the MaTYSSE[11] Large Programme (MaTYSSE stands for «MAgnetic Topologies of Young Stars and the Survival of close-in giant Exoplanets»), whereas ESPaDOnS / GRACES observations were collected in Director Discretionary Time in spectroscopic mode (no polarimetric unit on Gemini-N). All spectra were derived from raw frames with the reference pipeline implementing optimal extraction and RV correction from telluric lines[16], yielding a typical RV precision of 30 m/s rms[33].

### Deriving mean line profiles with least-squares deconvolution (LSD)

LSD[16] is a multiline technique similar to cross-correlation, used to derive line profiles with enhanced S/N from thousands of spectral lines simultaneously. For this study, the line list we used for LSD is derived from spectrum synthesis through model atmospheres computed assuming local thermodynamic equilibrium[34], for atmospheric parameters relevant for V830 Tau[12] (effective temperature of 4250 K, logarithmic gravity of 4.0 and solar metallicity). Resulting S/N in LSD profiles are in the range 950-1540 (see journal of observations, Extended Table 1), corresponding to average multiplex gains in S/N of ~10. LSD was also applied to circularly polarized spectra to retrieve average Zeeman signatures and longitudinal field estimates[16].

### Doppler imaging (DI) of stellar surfaces and the modeling of differential rotation

DI is a tomographic technique inspired from medical imaging, with which distributions of brightness features and magnetic fields at the surfaces of rotating stars can be reconstructed from time-series of high-resolution spectropolarimetric observations. DI is based on the fact that, thanks to the Doppler effect, line profiles of rotating stars can be interpreted as one-dimensional (1D) images of stellar surfaces, resolved in the Doppler direction but otherwise blurred. By coupling many such 1D images recorded at different rotation phases, one can reliably reconstruct the parent surface distribution that gives rise to the observed line profiles and rotational modulation. First introduced in the late 1980s[17], DI has been extensively used to investigate with unprecedented accuracy surface features and magnetic fields in cool stars other than the Sun[35,36]. Technically speaking, DI follows the principles of maximum-entropy image reconstruction, and iteratively looks for the image with lowest information content that fits the data at a given $\chi^2$ level. For more details on the imaging process, our previous MaTYSSE studies[11,12] give a detailed account of all modeling steps. In the case of the present paper, we carried out DI modeling using either unpolarized (Stokes I) spectra only, or both unpolarized and circularly polarized (Stokes V) spectra simultaneously, with identical results regarding filtering performances and RV results.

By looking at how surface maps get twisted as a function of time, DI can also estimate the amount of latitudinal differential rotation shearing stellar photospheres[18,19]. In this study we assume a typical solar-like differential rotation law in which the surface rotation rate varies with latitude $\theta$ as $\sin^2 \theta$, and depends on 2 main parameters, the rotation rate at the equator $\Omega_{eq}$ and the difference in rotation rate $d\Omega$ between the equator and the pole. Both parameters are derived by looking for the pair that minimizes the $\chi^2$ of the fit to the data (at constant information content in the reconstructed image, see Extended Fig 1), whereas the corresponding error bars are computed from the curvature of the $\chi^2$ paraboloid at its minimum[19]. Although helpful to achieve a more accurate description of the activity jitter and a cleaner filtering of raw RV curves (at periods $P_{rot}$ and $P_{rot}/2$ in particular), differential rotation as weak as that of V830 Tau has little impact on the filtered RVs; similar conclusions regarding the planet signal are obtained when assuming that V830 Tau is rotating as a solid body. A similar DI-based technique can be used to diagnose the presence of hJs around active stars[24], with the planet parameters replacing those describing differential rotation. This alternate method yields identical results to those presented here for our data set (see Fig 4).

**Filtering LSD profiles from lunar contamination**

Spectra recorded in non-photometric conditions near full-moon epochs are often contaminated by solar light reflected off the moon and diffused by clouds. To filter-out this pollution, whose location and width is well known at any given epoch but whose strength we want to determine, we implement a dual-step DI process. The first step consists in applying DI to the set of original LSD profiles, with scaled-up error bars for all pixels potentially affected by lunar contamination; the strength of the lunar contamination is then measured with a gaussian fit to the residuals, and subtracted from the polluted LSD profiles. In the second step, conventional DI is applied to the set of filtered LSD profiles with original error bars. This dual-step process is found to be very efficient when applied to densely-sampled data sets like the one presented here, in which profiles at similar phases but different rotation cycles provide a strong constraint on the strength of lunar pollution. In our data, 6 LSD profiles suffer from a strong pollution (rotation cycles 6.0 to 7.6, see Fig 1), whereas 9 others are affected at a much weaker level.

If excluding the 6 strongly (resp all 15) moon-polluted profiles from our data set, the RV signal from V830 Tau b is still clearly detected, though with a lower confidence rate of 99.9% (resp 99%) reflecting the poorer temporal coverage and the degraded window function (see Fig 3a and Extended Fig 2a). This check shows that our decontamination process is successful at restoring the original profile distortions and at retaining the RV content, provided the data set is dense enough.

**Revisiting the original data set from 2014 Dec & 2015 Jan**

A careful re-analysis of our original data (consisting of two subsets shifted in time by 17 d[12], see Extended Table 3) indicates that variability occurred at the surface of the star between the two subsets. While DI succeeds at adjusting the main subset (9 evenly-spaced points secured in 2015 Jan) down to noise level, as for our new data, fitting both subsets together requires to lower S/Ns in the 2014 Dec subset by ~15% in order to reach unit reduced $\chi^2$. This is definite evidence that intrinsic variability (beyond pure differential rotation) occurred at the surface of V830 Tau between both subsets, with profile #6 of the 2014 Dec subset being the most affected. This variability reflects a modification in the brightness map, subtle enough to affect DI no more than moderately, yet large enough to significantly affect filtered RVs, which are quite sensitive to even small features in the brightness map. It illustrates how tricky activity filtering can get when dealing with intrinsic variability, and how critical dense and even phase coverage is to efficiently diagnose it.

As a result of this variability, our filtering analysis can only be applied to the individual subsets of our original data, and in fact to no more than the 2015 Jan subset, the other being far too sparse and uneven for the technique to perform reliably. We find that the filtered RVs from the main subset (see Extended Table 3) agree with those of our new data; fitting them together improves the confidence level at which the planet is detected, but not the accuracy on the orbital period (see Extended Fig 2c).

Enabling to shortcut the computation of filtered RVs, the DI-based method of adjusting the planet parameters simultaneously with the distribution of surface features[24] offers an alternative option for confirming that the RV signal from the detected planet is present in our original data. (This method however still suffers from DI's inability to describe intrinsic variability beyond differential rotation). Freezing the planet orbital period to the value found in our new analysis (4.93 d) and applying this technique to the full set of our original data with only profile #6 removed, we find a semi-amplitude of 67±18 m/s for the planet RV signature, which agrees well with the measurement derived from our main study. We also confirm that our original data are better explained by a model including a planet in a 4.93-d orbit than by one with no planet, with a false-alarm probability of ~0.01% (corresponding to a $\chi^2$ increase of 19 for the 644 data points of the fitted LSD profiles).

**Code availability**

The Doppler imaging code used for this study is as yet undocumented and has thus not been released in the public domain.

**References con't**
31. Donati, J.-F. ESPaDOnS: An Echelle SpectroPolarimetric Device for the Observation of Stars at CFHT. *ASP Conf. Proc.* **307**, pp 41-50 (2003)
32. Chené, A.N., et al. GRACES: Gemini remote access to CFHT ESPaDOnS spectrograph through the longest astronomical fiber ever made: experimental phase completed, SPIE **9151**, 47-62 (2014)
33. Moutou, C., et al. Spectropolarimetric observations of the transiting planetary system of the K dwarf HD 189733, *Astron. & Astrophys.* **473**, 651-660 (2007)
34. Kurucz, R. ATLAS9 atmospheric models, ATLAS9 and SYNTHE routines, spectral line database. CDROM #13 & #18 (1993)
35. Collier Cameron, A. Spot Mapping in Cool Stars. *Lecture Notes in Physics* **573**, 183- (2001)
36. Morin, J. Magnetic Fields from Low-Mass Stars to Brown Dwarfs. *EAS Publ. Series* **57**, 165-191 (2012)

**Extended Data Legends**

**Extended Data Table 1 | Journal of observations**. Rotation and orbital cycles r and o are respectively given by the ephemeris BJD = 2,457,011.8 + 2.741 r and 2,457,360.52 + 4.93 o. For each observation, R, S/N and $S/N_{LSD}$ list the resolving power, the S/N in the raw spectrum and the S/N in the LSD profile, whereas the last 3 columns list the raw and filtered RVs and the corresponding 1σ error bars (reflecting both the photon noise and the instrumental RV precision).

**Extended Data Table 2 | Main parameters of the planet and of the host star.** Top table: main parameters (with 1σ error bars) of V830 Tau b assuming a circular orbit. Bottom table: main parameters (with 1σ error bars) of the host star V830 Tau[12].

**Extended Data Table 3 | Journal of observations for the original data.** Same as Extended Data Table 1 for the 2014 Dec and 2015 Jan data[12]. Our filtering technique is found to be only reliable for data sets with dense and regular phase coverage, hence the absence of filtered RVs for the 2014 Dec subset of 2x3 points, that does not satisfy this requirement. Filtered RVs from 2015 Jan 07-15 agree well with those derived from our new data (see Extended Data Table 1).

**Extended Data Figure 1 | Estimating surface differential rotation parameter.** Variations of the reduced $\chi^2$ as a function of the surface differential rotation parameters $\Omega_{eq}$ and $d\Omega$, denoting respectively the rotation rate at the equator and the difference in rotation rate between the equator and the pole (and assuming a solar-like sine-square differential rotation law). The location of the minimum and the local paraboloid curvature yield the optimal parameters and their respective 1σ error bars[19], equal to 2.29525±0.00020 and 0.0172±0.0014 rad/d. The outer color contour traces the 99.99% confidence interval (corresponding to a $\chi^2$ increase of 18.4 for a 2-parameter optimisation problem).

**Extended Data Figure 2 | Complementary periodograms. a**, Same as Fig 3a (middle plot) using the BGLS approach[21], showing that the 4.93-d peak we detect is at least $10^5$ time more likely than any other features. **b**, Same as Fig 3b for Hα emission, another activity proxy, featuring a clear peak at the stellar rotation period but no power at the planet orbital period. **c**, Same as Fig 3a (middle plot) for our new data combined with the 2015 Jan subsample of our original data[12]. The planet is now detected with a higher confidence level (FAP <$10^{-5}$) but the accuracy on the orbital period is not significantly improved (with multiple nearby peaks of similar strength). The red and green dashed lines are for the original and new data respectively. **d**, Same as Extended Data Figure 2c, zooming on the orbital frequency. The orbital period corresponding to the strongest peak is equal to 4.924±0.004 d (1σ error bar).

**Extended Data Figure 3 | Periodograms of simulated data. a**, Same as Fig 3, for simulated data computed using the brightness map, differential rotation and planet parameters inferred from the real data, and assuming the same coverage and similar S/N (equal for all LSD profiles). As for our observations, the planet signal is detected at a confidence level >99.9% in the filtered RVs despite being invisible in the raw RVs, and the planet parameters are well recovered. The periodogram of the raw RVs is very similar to that of Fig 3, featuring the main peaks (at $P_{rot}$ and $P_{rot}/2$) and their aliases; residual RVs mostly reflect the noise in the data. **b**, Same as Extended Data Figure 3a but with no planet included in the simulation. No signal with a confidence level >90% is recovered in the filtered RVs, demonstrating that the filtering process is not generating spurious RV signals, in particular at a period of 4.93 d.

**Extended Data Figure 4 | RV curves of simulated data. a**, Same as Fig 2a for the simulated data described in Extended Data Figure 3a (using the same brightness map, differential rotation and planet parameters as those of V830 Tau b). The simulated RVs (periodograms shown in Extended Data Figure 3a) share obvious similarities with the observed ones, and the planet signal is safely recovered (rms dispersion of residual RVs equal to 45 m/s). **b**, Same as **a** (middle panel) but with no planet included. As for the periodogram (see Extended Data Figure 3b, bottom panel), no signal is detected, further illustrating that activity induces no spurious planet signature. As for Fig 2, simulated RV measurements are depicted in all panels with their 1σ error bars.

# Figure 1

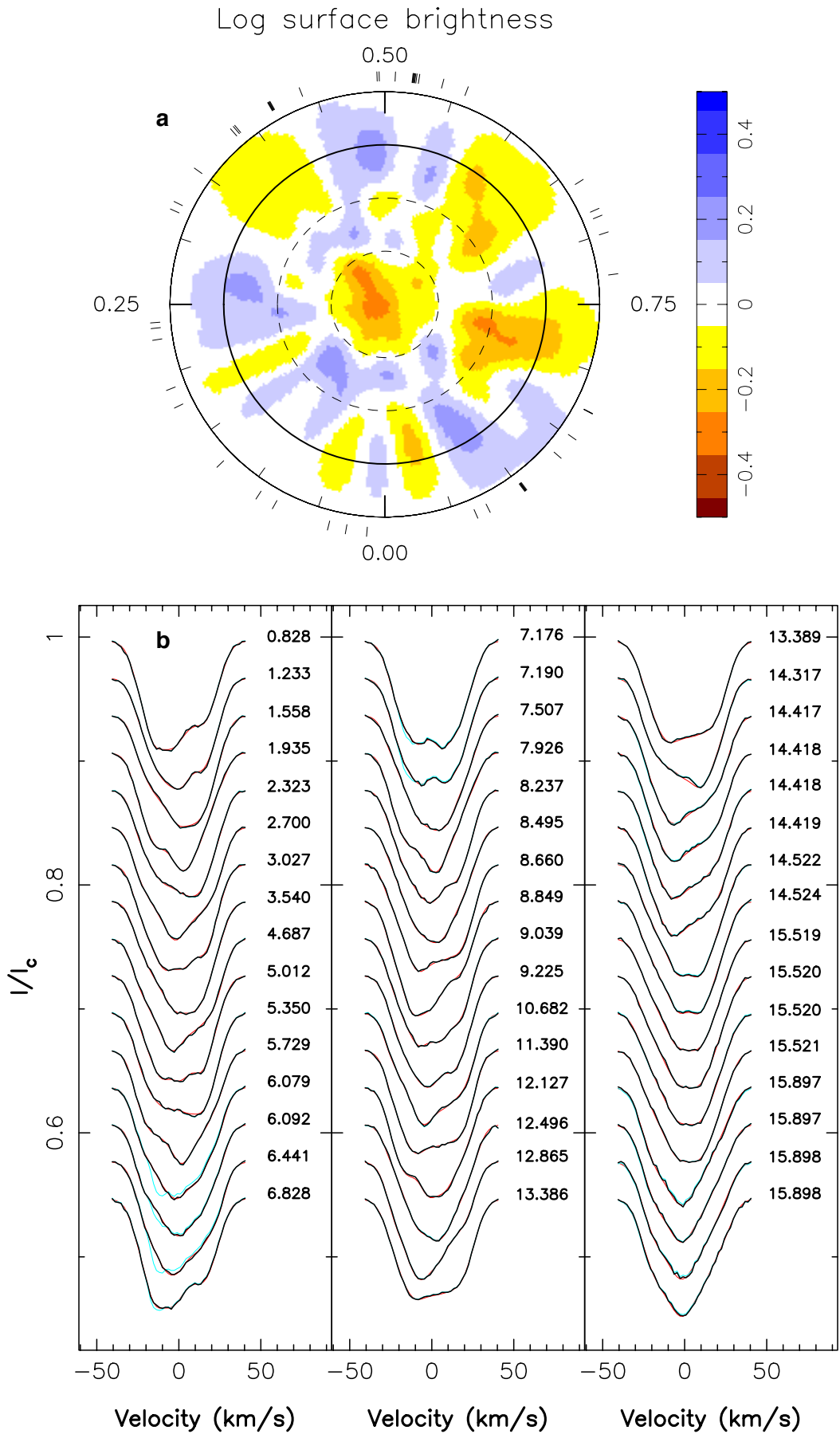

**Figure 2**

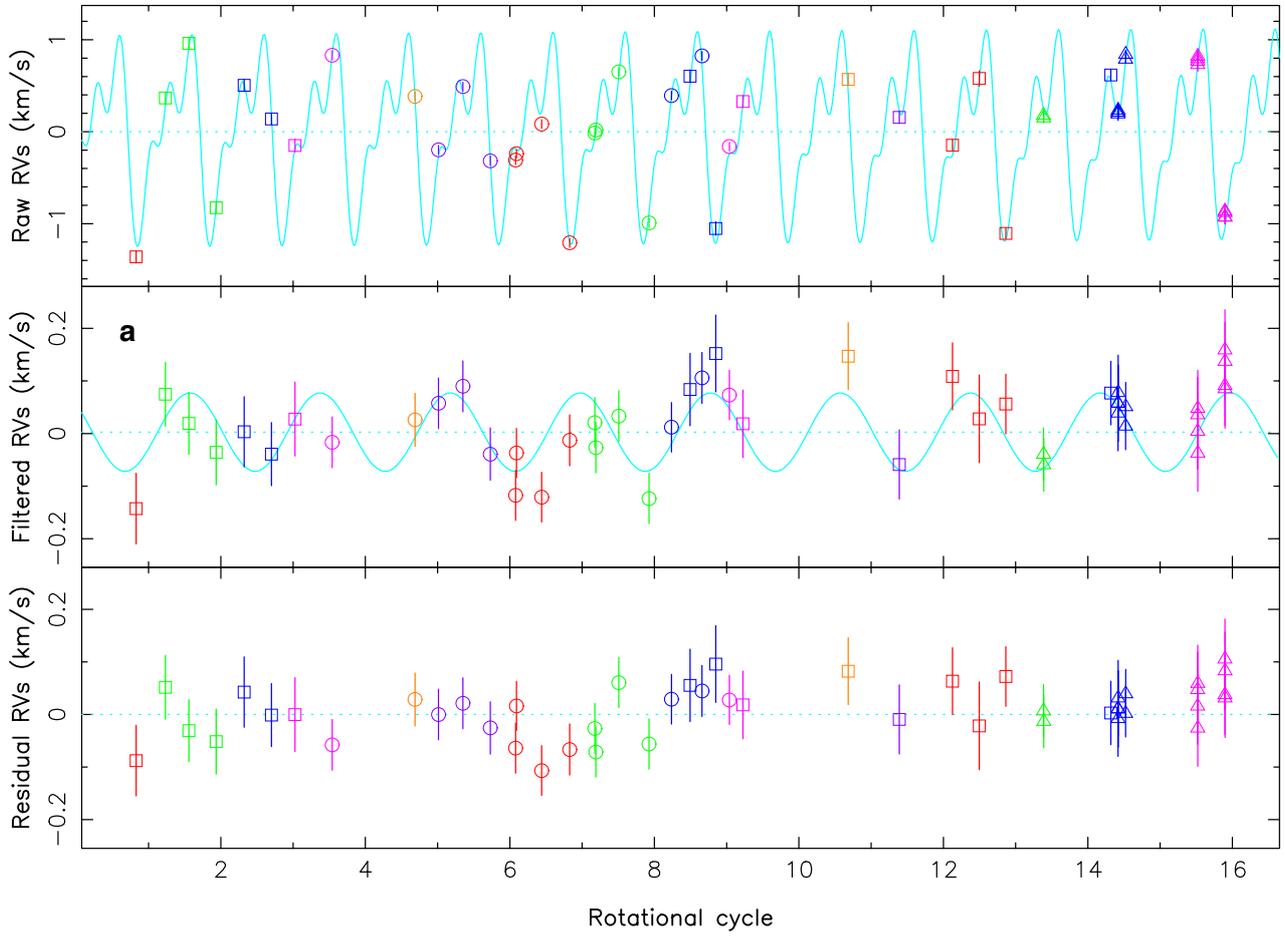

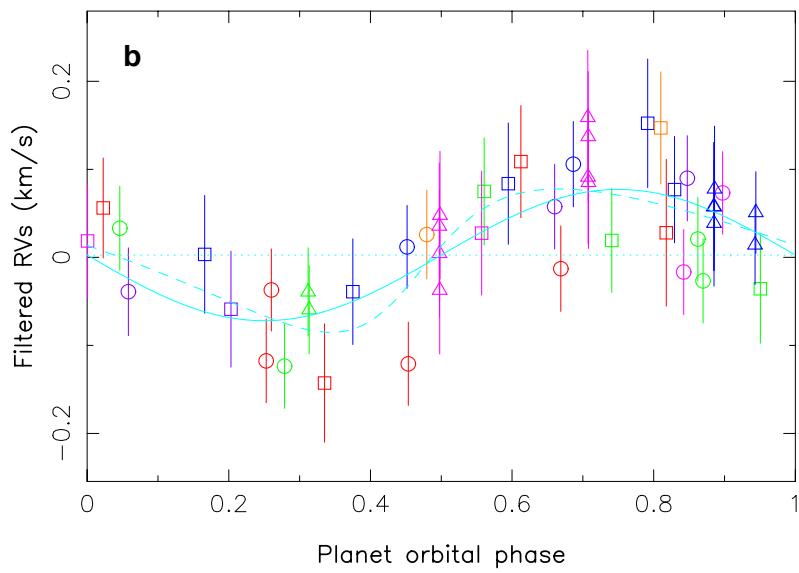

**Figure 3**

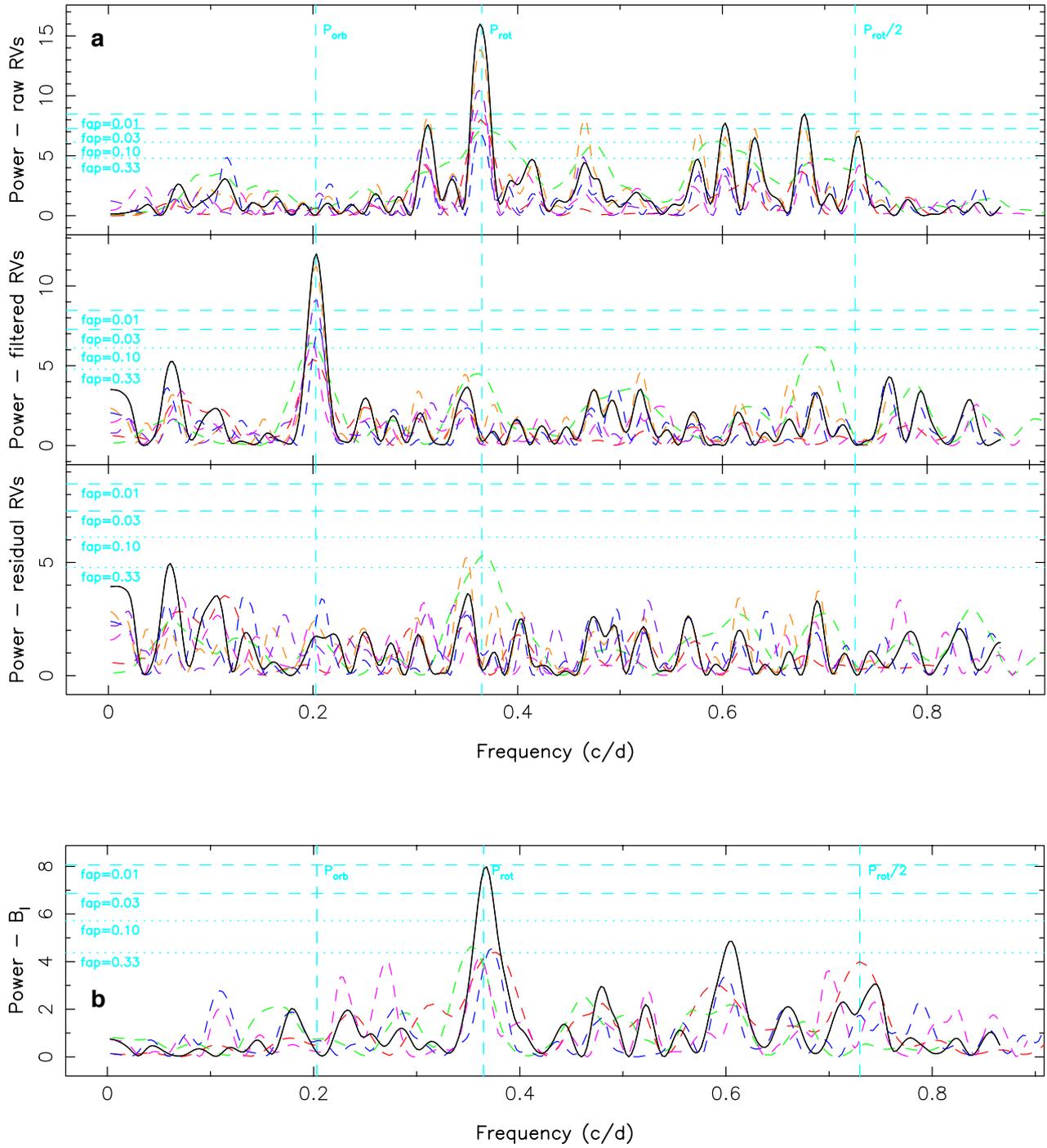

**Figure 4**

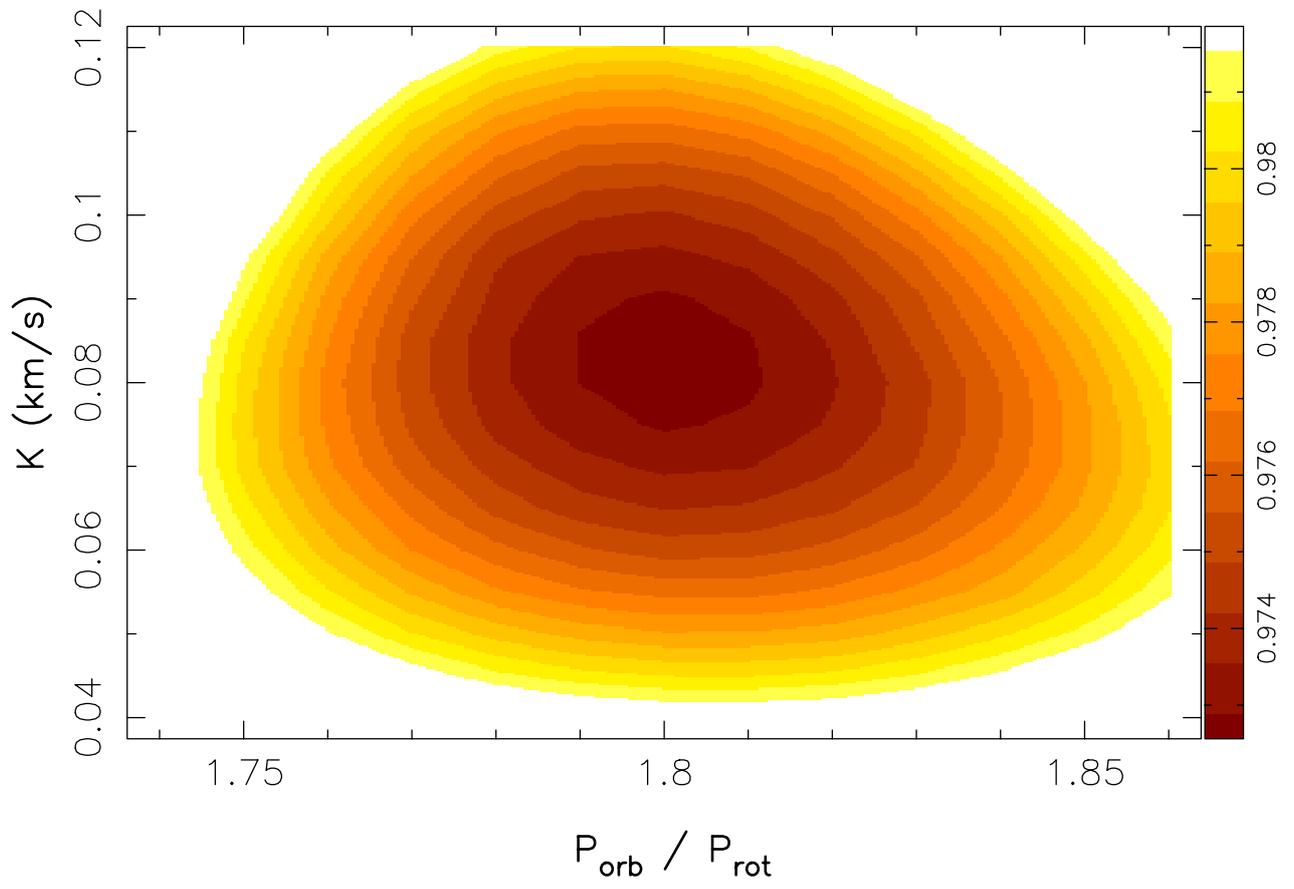

# Extended Data Table 1

| UT date (2015) | instrument | BJD (2457300+) | R (K) | $t_{exp}$ (s) | S/N | S/N$_{LSD}$ | rot cycle r (118+) | orb cycle o (5-) | raw RV (km/s) | filt RV (km/s) | RV err (km/s) |
|---|---|---|---|---|---|---|---|---|---|---|---|
| Nov 11 | NARVAL | 37.5066 | 65 | 4800 | 90 | 1149 | 0.828 | 0.332 | -1.362 | -0.150 | 0.067 |
| Nov 12 | NARVAL | 38.6187 | 65 | 4800 | 100 | 1251 | 1.233 | 0.557 | 0.365 | 0.074 | 0.061 |
| Nov 13 | NARVAL | 39.5077 | 65 | 4800 | 103 | 1278 | 1.558 | 0.738 | 0.960 | 0.022 | 0.059 |
| Nov 14 | NARVAL | 40.5427 | 65 | 4800 | 95 | 1233 | 1.935 | 0.948 | -0.828 | -0.035 | 0.062 |
| Nov 15 | NARVAL | 41.6051 | 65 | 4800 | 89 | 1152 | 2.323 | 1.163 | 0.504 | 0.005 | 0.067 |
| Nov 16 | NARVAL | 42.6375 | 65 | 4800 | 105 | 1257 | 2.700 | 1.373 | 0.137 | -0.036 | 0.060 |
| Nov 17 | NARVAL | 43.5347 | 65 | 4800 | 84 | 1113 | 3.027 | 1.555 | -0.151 | 0.031 | 0.070 |
| Nov 18 | ESPaDOnS | 44.9422 | 65 | 2780 | 162 | 1420 | 3.540 | 1.840 | 0.830 | -0.016 | 0.048 |
| Nov 21 | ESPaDOnS | 48.0862 | 65 | 2780 | 168 | 1380 | 4.687 | 2.478 | 0.382 | 0.031 | 0.050 |
| Nov 22 | ESPaDOnS | 48.9770 | 65 | 2780 | 170 | 1428 | 5.012 | 2.658 | -0.199 | 0.060 | 0.048 |
| Nov 23 | ESPaDOnS | 49.9011 | 65 | 2780 | 169 | 1428 | 5.350 | 2.846 | 0.489 | 0.087 | 0.048 |
| Nov 24 | ESPaDOnS | 50.9411 | 65 | 2780 | 158 | 1396 | 5.729 | 3.057 | -0.317 | -0.041 | 0.050 |
| Nov 25 | ESPaDOnS | 51.9015 | 65 | 2780 | 154 | 1482 | 6.079 | 3.252 | -0.312 | -0.120 | 0.048 |
| Nov 25 | ESPaDOnS | 51.9372 | 65 | 2780 | 164 | 1480 | 6.092 | 3.259 | -0.243 | -0.040 | 0.047 |
| Nov 26 | ESPaDOnS | 52.8917 | 65 | 2780 | 164 | 1539 | 6.441 | 3.452 | 0.089 | -0.126 | 0.047 |
| Nov 27 | ESPaDOnS | 53.9535 | 65 | 2780 | 160 | 1439 | 6.828 | 3.668 | -1.207 | -0.016 | 0.049 |
| Nov 28 | ESPaDOnS | 54.9089 | 65 | 2780 | 144 | 1458 | 7.177 | 3.862 | -0.014 | 0.018 | 0.047 |
| Nov 28 | ESPaDOnS | 54.9449 | 65 | 2780 | 131 | 1453 | 7.190 | 3.869 | 0.020 | -0.028 | 0.048 |
| Nov 29 | ESPaDOnS | 55.8153 | 65 | 2780 | 166 | 1434 | 7.507 | 4.045 | 0.648 | 0.029 | 0.048 |
| Nov 30 | ESPaDOnS | 56.9632 | 65 | 2780 | 175 | 1436 | 7.926 | 4.278 | -0.992 | -0.123 | 0.048 |
| Dec 01 | ESPaDOnS | 57.8164 | 65 | 2780 | 167 | 1447 | 8.237 | 4.451 | 0.393 | 0.014 | 0.047 |
| Dec 02 | NARVAL | 58.5216 | 65 | 4800 | 89 | 1131 | 8.495 | 4.594 | 0.602 | 0.078 | 0.069 |
| Dec 02 | ESPaDOnS | 58.9744 | 65 | 2780 | 168 | 1428 | 8.660 | 4.686 | 0.823 | 0.113 | 0.049 |
| Dec 02 | NARVAL | 59.4928 | 65 | 4800 | 82 | 1089 | 8.849 | 4.791 | -1.054 | 0.148 | 0.073 |
| Dec 03 | ESPaDOnS | 60.0148 | 65 | 2780 | 176 | 1458 | 9.039 | 4.897 | -0.163 | 0.073 | 0.047 |
| Dec 04 | NARVAL | 60.5247 | 65 | 4800 | 90 | 1191 | 9.225 | 5.001 | 0.328 | 0.019 | 0.064 |
| Dec 07 | NARVAL | 64.5183 | 65 | 4800 | 108 | 1221 | 10.682 | 5.811 | 0.568 | 0.151 | 0.064 |
| Dec 09 | NARVAL | 66.4571 | 65 | 4800 | 89 | 1161 | 11.390 | 6.204 | 0.155 | -0.068 | 0.066 |
| Dec 11 | NARVAL | 68.4771 | 65 | 4800 | 88 | 1218 | 12.127 | 6.614 | -0.148 | 0.103 | 0.064 |
| Dec 12 | NARVAL | 69.4897 | 65 | 4800 | 69 | 978 | 12.496 | 6.819 | 0.578 | 0.024 | 0.083 |
| Dec 13 | NARVAL | 70.5011 | 65 | 4800 | 104 | 1335 | 12.865 | 7.024 | -1.107 | 0.055 | 0.057 |
| Dec 15 | ESPaDOnS / GRACES | 71.9290 | 35 | 540 | 218 | 1385 | 13.386 | 7.314 | 0.178 | -0.048 | 0.068 |
| Dec 15 | ESPaDOnS / GRACES | 71.9359 | 35 | 540 | 214 | 1388 | 13.389 | 7.315 | 0.148 | -0.068 | 0.068 |
| Dec 17 | NARVAL | 74.4815 | 65 | 4800 | 90 | 1257 | 14.317 | 7.832 | 0.617 | 0.080 | 0.060 |
| Dec 18 | ESPaDOnS / GRACES | 74.7548 | 35 | 85 | 72 | 974 | 14.417 | 7.887 | 0.222 | 0.058 | 0.073 |
| Dec 18 | ESPaDOnS / GRACES | 74.7565 | 35 | 85 | 73 | 992 | 14.418 | 7.888 | 0.203 | 0.039 | 0.072 |
| Dec 18 | ESPaDOnS / GRACES | 74.7581 | 35 | 85 | 78 | 990 | 14.418 | 7.888 | 0.222 | 0.057 | 0.072 |
| Dec 18 | ESPaDOnS / GRACES | 74.7597 | 35 | 85 | 77 | 996 | 14.419 | 7.888 | 0.242 | 0.078 | 0.071 |
| Dec 18 | ESPaDOnS / GRACES | 75.0425 | 65 | 360 | 123 | 1396 | 14.522 | 7.946 | 0.788 | 0.016 | 0.045 |
| Dec 18 | ESPaDOnS / GRACES | 75.0474 | 65 | 360 | 107 | 1369 | 14.524 | 7.947 | 0.840 | 0.053 | 0.046 |
| Dec 21 | ESPaDOnS / GRACES | 77.7761 | 35 | 85 | 75 | 990 | 15.519 | 8.500 | 0.792 | 0.037 | 0.072 |
| Dec 21 | ESPaDOnS / GRACES | 77.7777 | 35 | 85 | 74 | 986 | 15.520 | 8.500 | 0.766 | 0.006 | 0.072 |
| Dec 21 | ESPaDOnS / GRACES | 77.7793 | 35 | 85 | 74 | 982 | 15.520 | 8.501 | 0.729 | -0.035 | 0.073 |
| Dec 21 | ESPaDOnS / GRACES | 77.7810 | 35 | 85 | 74 | 976 | 15.521 | 8.501 | 0.819 | 0.050 | 0.073 |
| Dec 22 | ESPaDOnS / GRACES | 78.8110 | 35 | 85 | 58 | 946 | 15.897 | 8.710 | -0.864 | 0.161 | 0.076 |
| Dec 22 | ESPaDOnS / GRACES | 78.8126 | 35 | 85 | 64 | 956 | 15.897 | 8.710 | -0.928 | 0.093 | 0.076 |
| Dec 22 | ESPaDOnS / GRACES | 78.8143 | 35 | 85 | 61 | 964 | 15.898 | 8.711 | -0.880 | 0.139 | 0.074 |
| Dec 22 | ESPaDOnS / GRACES | 78.8159 | 35 | 85 | 62 | 958 | 15.899 | 8.711 | -0.929 | 0.087 | 0.075 |

# Extended Data Table 2

| orbital period (d) | K (m/s) | BJD of transit | orbital distance (au) | $m_{planet} \sin i$ ($M_{2\!\!\downarrow}$) | $m_{planet}$ ($M_{2\!\!\downarrow}$) |
|---|---|---|---|---|---|
| 4.93±0.05 | 75±11 | 2457360.52±0.10 | 0.057±0.001 | 0.63±0.11 | 0.77±0.15 |

| $M_{star}$ ($M_\odot$) | $R_{star}$ ($R_\odot$) | age (Myr) | $T_{eff}$ (K) | $\log(L/L_\odot)$ | $P_{rot}$ (d) | $v \sin i$ (km/s) | i (°) | distance (pc) |
|---|---|---|---|---|---|---|---|---|
| 1.00±0.05 | 2.0±0.2 | ~2 | 4250±50 | 0.08±0.10 | 2.741 | 30.5±0.5 | 55±10 | 131±3 |

# Extended Data Table 3

| UT date (2014-2015) | instrument | BJD (2457000+) | R (K) | $t_{exp}$ (s) | S/N | S/N$_{LSD}$ | rot cycle r | raw RV (km/s) | filt RV (km/s) | RV err (km/s) |
|---|---|---|---|---|---|---|---|---|---|---|
| Dec 20 | ESPaDOnS | 11.8899 | 65 | 2800 | 170 | 1501 | 0.033 | 0.721 |  | 0.049 |
| Dec 21 | ESPaDOnS | 12.8622 | 65 | 2800 | 170 | 1504 | 0.388 | 0.197 |  | 0.050 |
| Dec 22 | ESPaDOnS | 13.9010 | 65 | 2800 | 180 | 1520 | 0.767 | -0.513 |  | 0.049 |
| Dec 28 | ESPaDOnS | 20.0190 | 65 | 2800 | 140 | 1498 | 2.999 | 0.446 |  | 0.049 |
| Dec 29 | ESPaDOnS | 20.8759 | 65 | 2800 | 160 | 1478 | 3.311 | -0.137 |  | 0.050 |
| Dec 30 | ESPaDOnS | 21.8154 | 65 | 2800 | 160 | 1478 | 3.654 | 0.240 |  | 0.050 |
| Jan 07 | ESPaDOnS | 29.8629 | 65 | 2800 | 170 | 1498 | 6.590 | -0.044 | 0.031 | 0.049 |
| Jan 08 | ESPaDOnS | 30.8217 | 65 | 2800 | 180 | 1490 | 6.940 | -0.206 | -0.031 | 0.049 |
| Jan 09 | ESPaDOnS | 31.8181 | 65 | 2800 | 170 | 1523 | 7.303 | -0.151 | -0.062 | 0.051 |
| Jan 10 | ESPaDOnS | 32.8186 | 65 | 2800 | 150 | 1476 | 7.668 | 0.044 | -0.011 | 0.050 |
| Jan 11 | ESPaDOnS | 33.8669 | 65 | 2800 | 180 | 1495 | 8.051 | 1.085 | 0.076 | 0.050 |
| Jan 12 | ESPaDOnS | 34.7215 | 65 | 2800 | 170 | 1473 | 8.362 | 0.246 | 0.073 | 0.050 |
| Jan 13 | ESPaDOnS | 35.7150 | 65 | 2800 | 160 | 1501 | 8.725 | -0.264 | -0.001 | 0.050 |
| Jan 14 | ESPaDOnS | 36.7141 | 65 | 2800 | 170 | 1501 | 9.089 | 0.929 | -0.067 | 0.050 |
| Jan 15 | ESPaDOnS | 37.8043 | 65 | 2800 | 170 | 1470 | 9.487 | -0.202 | -0.052 | 0.050 |

**Extended Data Figure 1**

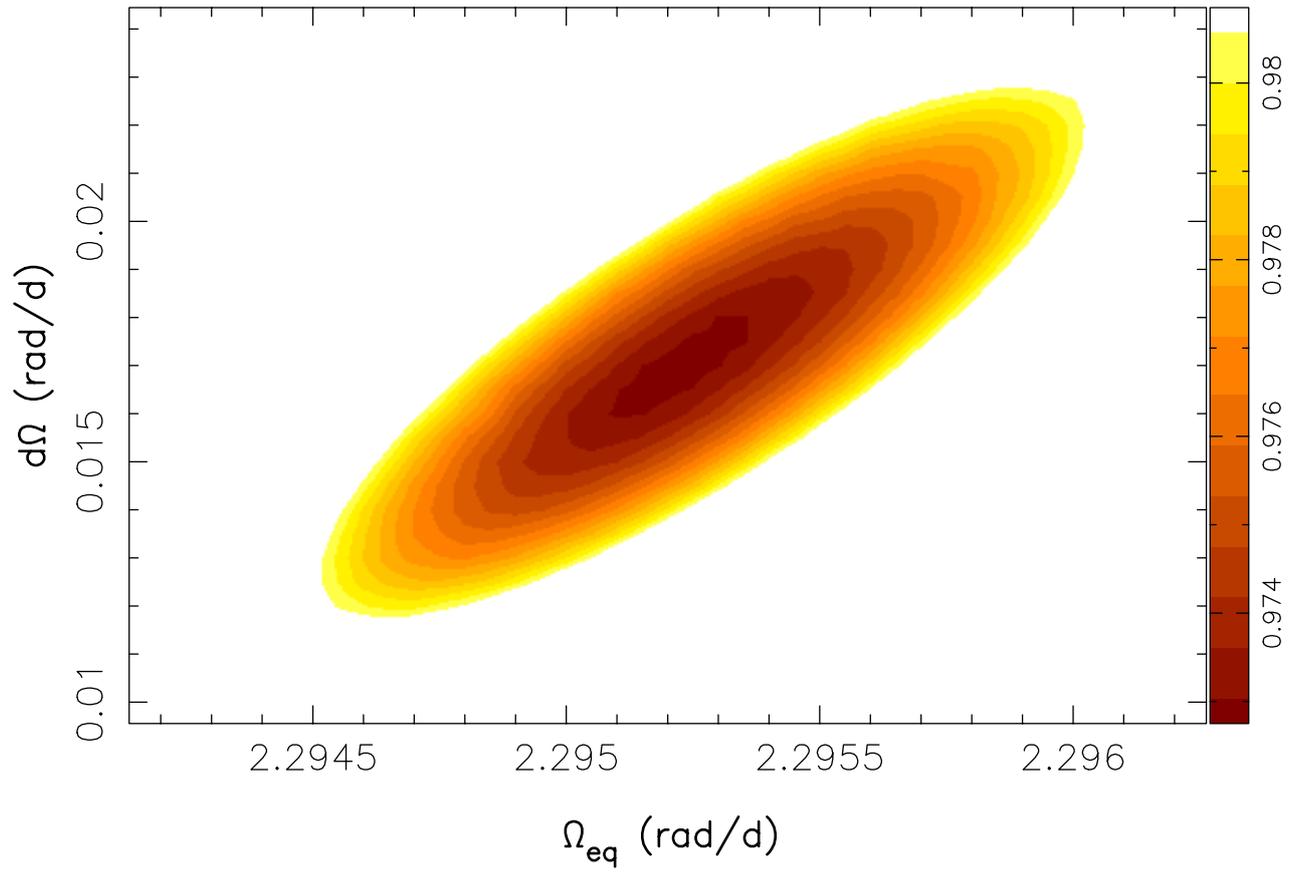

# Extended Data Figure 2

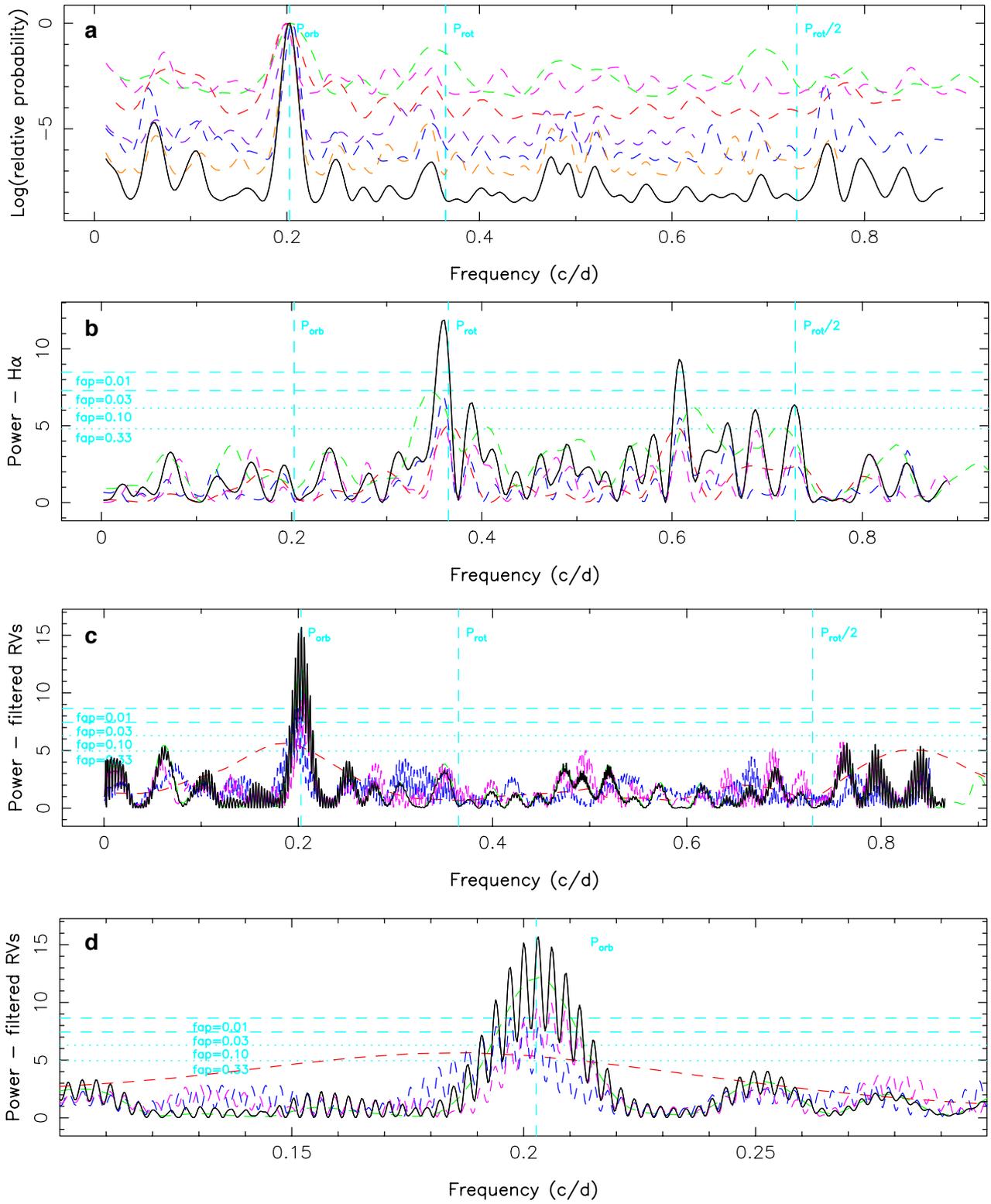

**Extended Data Figure 3**

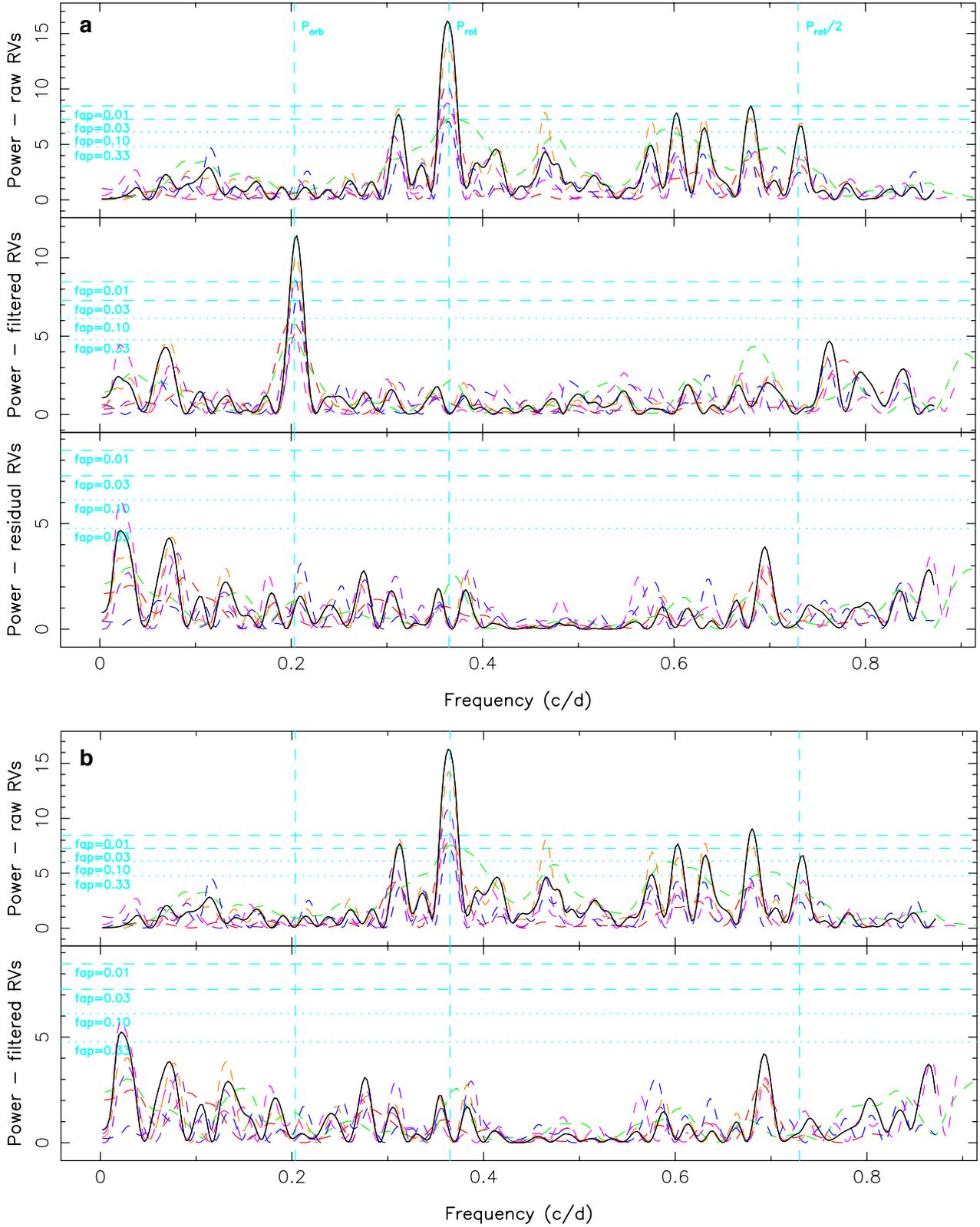

**Extended Data Figure 4**

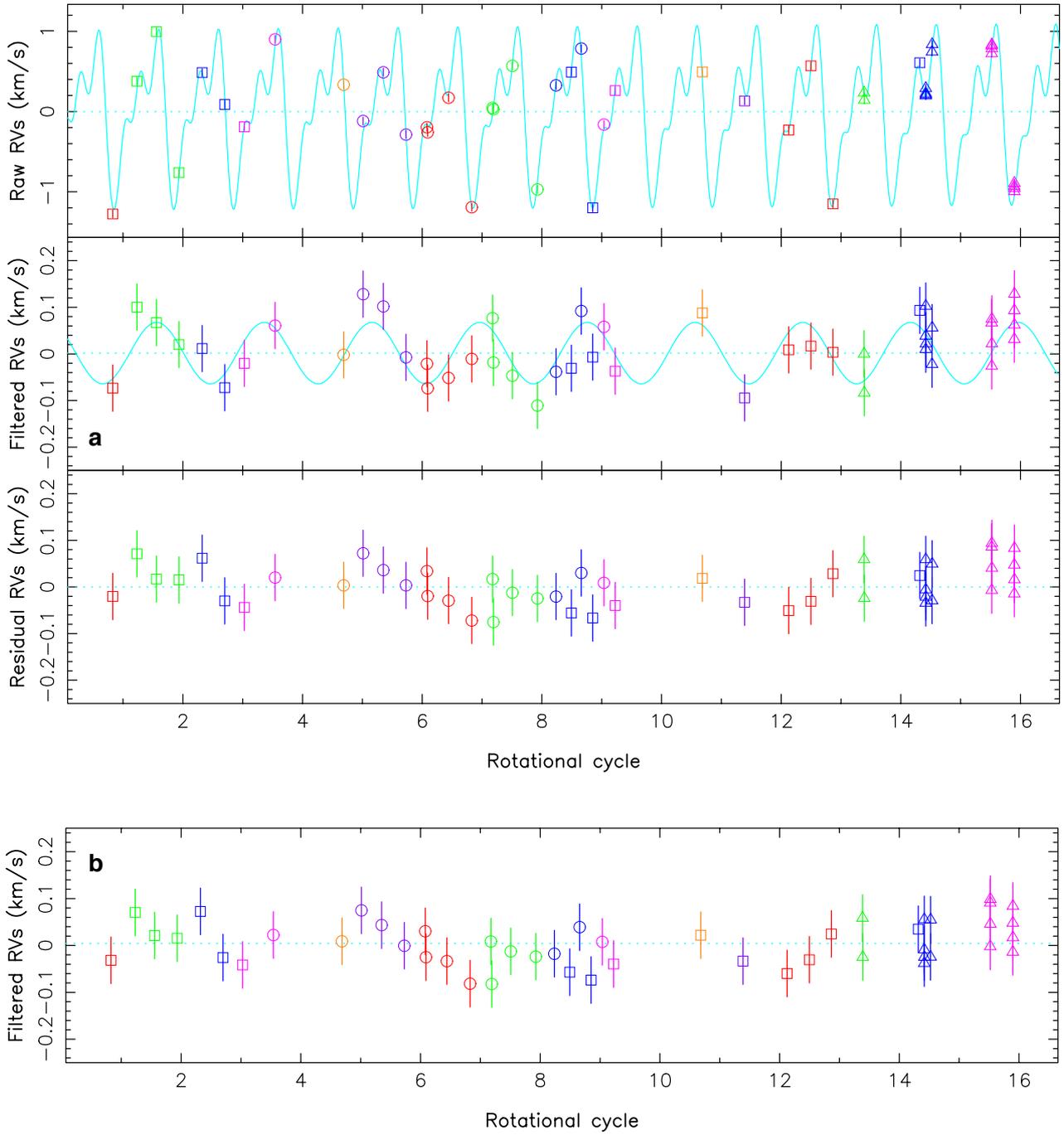